\documentclass{article}

\usepackage{arxiv}

\usepackage[utf8]{inputenc} 
\usepackage[english]{babel}
\addto\captionsenglish{%
}
\usepackage[T1]{fontenc}    
\usepackage {xcolor}
\usepackage{hyperref}
\hypersetup{
    colorlinks = true,
    linkbordercolor = {white},
    citecolor    = gray,
    urlcolor     = blue
}
\usepackage{url}            
\usepackage{booktabs}       
\usepackage{amsmath}        %
\usepackage{amsfonts}       
\usepackage{nicefrac}       
\usepackage{microtype}      
\usepackage{graphicx}
\usepackage{lipsum}		
\usepackage[toc,page]{appendix}
\usepackage{lineno}
\modulolinenumbers[5]
\usepackage{cclicenses}
\usepackage{makecell}
\usepackage{lscape,array}
\newcolumntype{C}[1]{>{\centering\arraybackslash}p{#1}} 
\usepackage[thin, , thinc]{esdiff}
\usepackage{subcaption}
\usepackage{caption}
\usepackage{framed}  
\usepackage[font=small,skip=1pt]{caption}
\usepackage{longtable}
\usepackage{soul}
\graphicspath{{images/}} 

\title{Universal and non-universal text statistics: Clustering coefficient for language identification}

\author{
 Diego Espitia \\
  Centro de Investigaciones en Ciencias,\\
  Universidad Autónma del Estado de Morelos, and\\
  Instituto de Ciencias Físicas\\
  Universidad Nacional Auntónoma de México\\
  \texttt{despitia@icf.unam.mx} \\
  \and
    Hernán Larralde Ridaura \\
  Instituto de Ciencias Físicas\\
  Universidad Nacional Auntónoma de México\\
  \texttt{hernan@icf.unam.mx} \\
  }

\begin{document}
\maketitle

\begin{abstract}
In this work we analyze statistical properties of 91 relatively small texts in 7 different languages (Spanish, English, French, German, Turkish, Russian, Icelandic) as well as texts with randomly inserted spaces. Despite the size (around 11260 different words), the well known universal statistical laws -namely Zipf and Herdan-Heap's laws- are confirmed, and are in close agreement with results obtained elsewhere. 
We also construct a word co-occurrence network of each text. While the degree distribution is again universal, we note that the distribution of clustering coefficients, which depend strongly on the local structure of networks, can be used to differentiate between languages, as well as to distinguish natural languages from random texts.
\end{abstract}


\section{Introduction}

Statistical characterization of languages has been a field of study for 
decades\cite{Zipf,GellMann,Konto,Kohler,Zanette,AltmannStat}. Even simple quantities, 
like letter frequency, can be used to decode simple substitution 
cryptograms\cite{Voynich,VoyAmancio,VoyMontemurro,Copiale,Copiale2}. However, probably 
the most surprising result in the field is Zipf's law, which states that if one ranks 
words by their frequency in a large text, the resulting rank 
frequency distribution is approximately a power law, for all languages 
\cite{Zipf,Piantadosi2014}. These kind of universal results have long piqued the 
interest of physicists and mathematicians, as well as 
linguists\cite{Zipf4Phrases,ZipfEspanoles,Newman2000}. Indeed, a large 
amount of effort has been devoted to try to understand the origin of 
Zipf's law, in some cases arguing that it arises from the fact that 
texts carry information \cite{FerreriCancho2005}, all the way to arguing that it is the 
result of mere chance \cite{ZornigRandomZipf,WentianLi}. Another interesting 
characterization of texts is the Heaps-Herdan law, which describes how the vocabulary 
-that is, the set of different words- grows with the size of a text, the number of 
which, empirically, has been found to grow as a power of the 
text size \cite{HerdanBook,HeapsBook}. It is worth noting that it 
has been argued that this law is a consequence Zipf's law. \cite{HeapsBaeza,egghe_2007}

A different tool used to characterize texts is the adjacency (or co-ocurrence) network 
\cite{CanchoRedes,AmancioRedes,Liu2013,JaposKeyword}. The nodes in this network 
represent the words in the text, and a link is placed between nodes if the corresponding
words are adjacent in the text. These links can be directed -according to the order in 
which the words appear-, or undirected. In this work we study properties of the 
adjacency network of various texts in several languages, using undirected links.  
The advantage of representing the text as a network is that we can 
describe properties of the text using the tools of network theory 
 \cite{BarabasiBook}. The simplest characterization of a network is its degree 
distribution, that is, the fraction of nodes with a given number of 
links, and we will see that this distribution is also a universal power 
law for all languages. As we argue ahead, this may follow from the fact 
that Zipf's law is satisfied. 

Another interesting use for text statistics is to distinguish texts and 
languages. In particular, as occurs with letter frequencies, other more 
subtle statistics may be used to distinguish different languages, and 
beyond that, provide a metric to group languages into different families 
\cite{Gamallo,Barry,PETRONI2010}. In this paper we use the clustering coefficient 
\cite{BarabasiBook} to show that even though the degree distribution of the adjacency 
matrices is common to all languages, the statistics of their clustering coefficients, 
while approximately similar for various texts in each language, appears to be different 
from one language to another.

We use different texts (see Appendix (\ref{Libros})) instead of a large single corpus for 
each language because clustering coefficients typically decrease as a function of the 
size of the network\cite{MeanFieldClustering}. Actually, we must compare the statistics 
of the clustering coefficient in texts with adjacency networks of comparable sizes. In 
the following section we present the rank vs frequency distribution for these 
texts. We also measure how the vocabulary increases with text size, as well as the 
respective degree distributions of the networks corresponding to every text, and compare
them with a null "random" hypothesis. This null hypothesis consists of a set of texts 
constructed as follows: we select a text and remove all the spaces between words, then 
we reintroduce the spaces at random with the restriction that there cannot be a space 
next to another. We identify as words all strings of letters between consecutive spaces 
(the restriction avoids the possibility of having empty words). The reason we build the 
null hypothesis this way instead of the usual independent random letters with random 
spaces most commonly used \cite{WentianLi,Miller}, is that consecutive letters are not independent: they 
are correlated to ensure word pronunciability, as well as due to spelling rules. Our 
method for constructing these random texts conserves most of the correlations between 
consecutive letters in a given language.

Next, we calculate the distribution of the clustering coefficients of the nodes of the 
adjacency network for each text. These distribution functions are more or less similar 
for all the texts of the same language, provided the networks are of the same size. 
However, it is apparent that the distributions are different between 
different languages. We also compare the clustering coefficient 
distributions with those of the null hypothesis. The data show that the strongest 
differences between languages occur for the fractions of nodes with clustering 
coefficients 0 and 1. We build a scatter plot for these fractions for all the texts in 
each language. Though there is overlap between some languages, other languages are 
clearly differentiated in the plot. We fit correlated bivariate gaussian distributions 
to the data of each language, which allows us to estimate a likelihood that a text is in a given language.

\section{Texts and Universal laws}

We analyzed 91 texts written in 7 languages: Spanish, English, German, French, Turkish, 
Russian and Icelandic. We also considered as null texts, 12 realizations of a randomized
version the Portrait of Dorian Gray book, twice for each language analyzed here (except 
Icelandic). As mentioned above, the process for randomizing the text is as follows: first we remove the spaces in the original text. Then, we take the first letter, and with a probability of $1/2$ we add the next letter in the sequence, or the next letter in the sequence and a space. We advance to the last symbol added, and repeat the process until we reach the end of the text. This way we destroy the grammar of the original language, keeping the letter frequencies as well as most of the correlations between consecutive letters.
The set of documents we used in this work are shown in Appendix \ref{Libros}.

All texts were intervened to remove punctuation marks, 
numbers, parenthesis and other uncommon symbols, and all the letters 
were turned into lower case, so a word appearing with different case 
letters would not be counted as two different words. Also, we do not transliterate the texts, instead, we use the original symbols of the texts (Cyrillic alphabet for Russian texts or the special characters in Icelandic) using the UTF-8 encoding. 

Also, since clustering coefficients depend non trivially on the size of the networks, we cut the texts so they all have essentially the same vocabulary size ($\simeq 11260$). 

In table \ref{Resumen} we summarize for each language, the averages of the length, vocabulary size, maximum frequency and number of hapax legomena (i.e. words that appear only once in a document or corpus) of the texts studied here. It is important to note that for different languages, very different text lengths are required to achieve the same vocabulary size. We also note that in all cases, hapax legomena represent approximately half of the vocabulary in each text.

\begin{table}[htb!]
\centering
\begin{tabular}{@{}lcccc@{}}
\toprule
Language  & Length & Vocabulary & Maximum Frequency & Number of Hapax \\ 
& $L$ & $N_{tot}$ & $f_{max}$ & $H$ 
\\ \midrule
Spanish   & 85920  & 11253      & 4460    & 6265                     \\
English   & 219539 & 11258      & 11961   & 4582                     \\
French    & 90797  & 11260      & 4001    & 6302                     \\
German    & 78464  & 11264      & 2724    & 6590                     \\
Turkish   & 35392  & 11246      & 1173    & 7593                     \\
Russian   & 43799  & 11271      & 1726    & 7541                     \\
Icelandic & 93699  & 11254      & 5063    & 6598                     \\
Random    & 63753  & 11258      & 4010    & 8340                     \\ \bottomrule \\
\end{tabular}
\caption{Averages for each language of the length, vocabulary size, maximum word frequency and number of hapax legomena of the texts studied in this work. Notice the large variations in text length required to achieve the same vocabulary size $N_{tot}$ from one language to another.}
\label{Resumen}
\end{table}

\begin{figure}[h]
\centering
\includegraphics[scale=0.7]{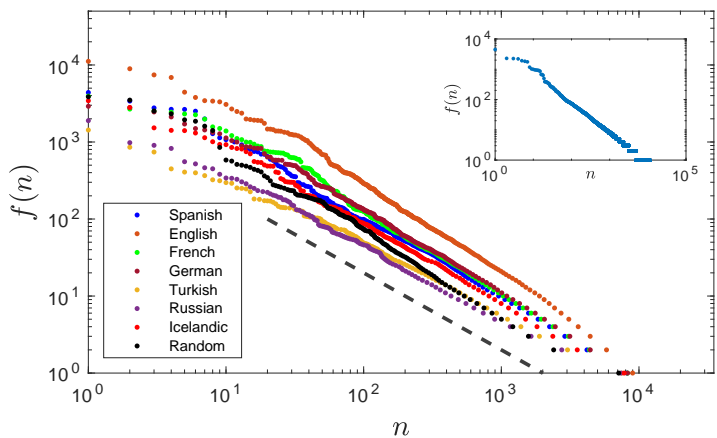}
\caption{\emph{Word frequency $f(n)$ versus rank $n$ illustrating Zipf's Law $f(n)\sim 1/n^\alpha$ for single randomly chosen texts in each language: English, Russian, Turkish, French, German, Spanish and Icelandic, using Log-binned data (colored symbols). Black dots represent the random texts constructed as described in the text. The dashed line corresponds to $\alpha=1$. In the inset we show an example of rank vs frequency plot without log-binning. Note that words with $f=1$ (hapax legomena) represent a large fraction of the vocabulary of the text}}
  \label{ZipfLaw}
\end{figure}

In figure (\ref{ZipfLaw}) we show Zipf plots for some of the texts, 
including the random texts constructed as described previously. It is clear that 
all the texts reproduce convincingly Zipf's law: $f(n)\sim 1/n^\alpha$ where $n=1,2,...N_{tot}$ is the word rank, $N_{tot}$ 
is the size of the vocabulary and $f(n)$ is its frequency. This is in contrast to previous work in which it is argued that there are differences between the Zipf 
plots of texts and random sequences\cite{ZipfCancho}, this might be due to the fact that our random text construction preserves correlations between letters, whereas the letters in \cite{ZipfCancho} were placed independently.  Our findings are summarized in Appendix (\ref{Resultados}).\footnote{We are aware that it has been argued that Zipf's law -namely a pure power law relation between rank and frequency- is not valid throughout the complete distribution, \cite{ZipfTwo, NAUMIS200884}. In this work we refer to Zipf's law as the power law behavior of the "tail" of the distribution (which comprises over 99\% of the vocabulary), and is also the region for which the Maximum Likelihood Estimator (MLE) method described in Appendix \ref{Resultados} is best suited for.}

\newpage 

Figure (\ref{ZipfLaw}) is the typical rank vs frequency plot for a randomly chosen text in each language. From the figure, we see that $\alpha \simeq 1$, obtained by least squares fits to the plot, describes very well all the texts. Therefore, 
given that $n/N_{tot}$ is the fraction of 
words with frequencies greater or equal to $f(n)$, then
\begin{equation}
\frac{n}{N_{tot}}\sim \int\limits_{f(n)}^\infty p(f)~df,
\label{eq.1}
\end{equation}

where $p(f) \simeq 1/f^{\alpha_z}$ is the frequency distribution of the vocabulary. Now, if $f(n)\sim 1/n^\alpha$, then $p(f)\sim 1/f^{1+1/\alpha}$, i.e. $\alpha_z=1+1/\alpha$. Substituting $\alpha=1$, we have $\alpha_Z = 2$, which is in close agreement with what we observe. See figure (\ref{CumP(f)})  and the tables in Appendix (\ref{Resultados})

\begin{figure}[h]
\centering
\includegraphics[scale=0.7]{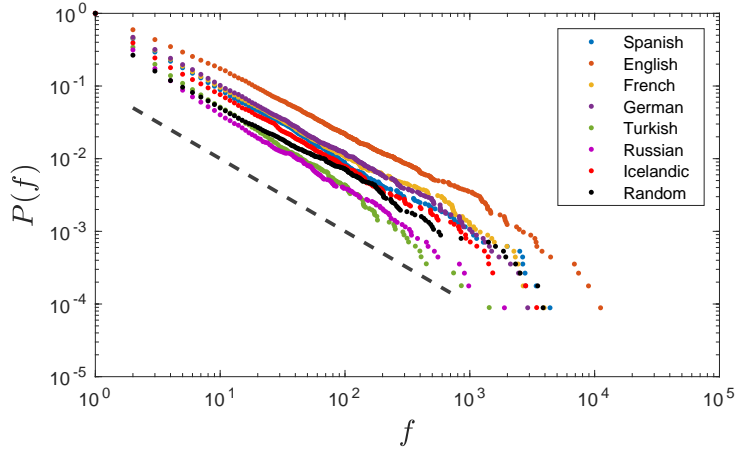}
\caption{\emph{Cumulative of the frequency distribution $P(f)\equiv\int_f^\infty p(\zeta) d\zeta$ for single randomly chosen texts in each language: Spanish, English, French, German, Turkish, Russian, and Icelandic (colored symbols). Black dots represent a random text, and the dashed line corresponds to the expected behavior when the exponent in Zipf's law is $\alpha=1$.}}
  \label{CumP(f)}
\end{figure}

\begin{figure}[h]
\centering
\includegraphics[scale=0.7]{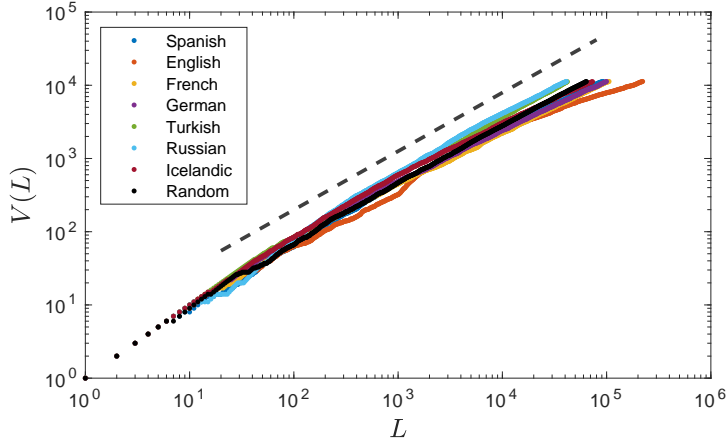}
\caption{\emph{The Herdan-Heap's Law $V(L)\sim  L^{\beta}$ for single randomly chosen texts in each language: English, Russian, French, German, Spanish, Icelandic and Turkish. Black dots represent random texts. The dashed line corresponds to a power law with exponent $\beta=0.8$}, which is the average over all the texts we studied.}
  \label{H&HLaw}
\end{figure}

Figure (\ref{H&HLaw}) shows the size of the vocabulary $V(L)$, as a function of the length $L$ of the text considered. Once again, all the texts, including the random texts, follow the Heaps-Herdan law $V(L)\sim  L^{\beta}$ reasonably well. Again, the parameters describing the various texts are given in Appendix(\ref{Resultados})

\begin{figure}[h]
\centering
\includegraphics[scale=0.7]{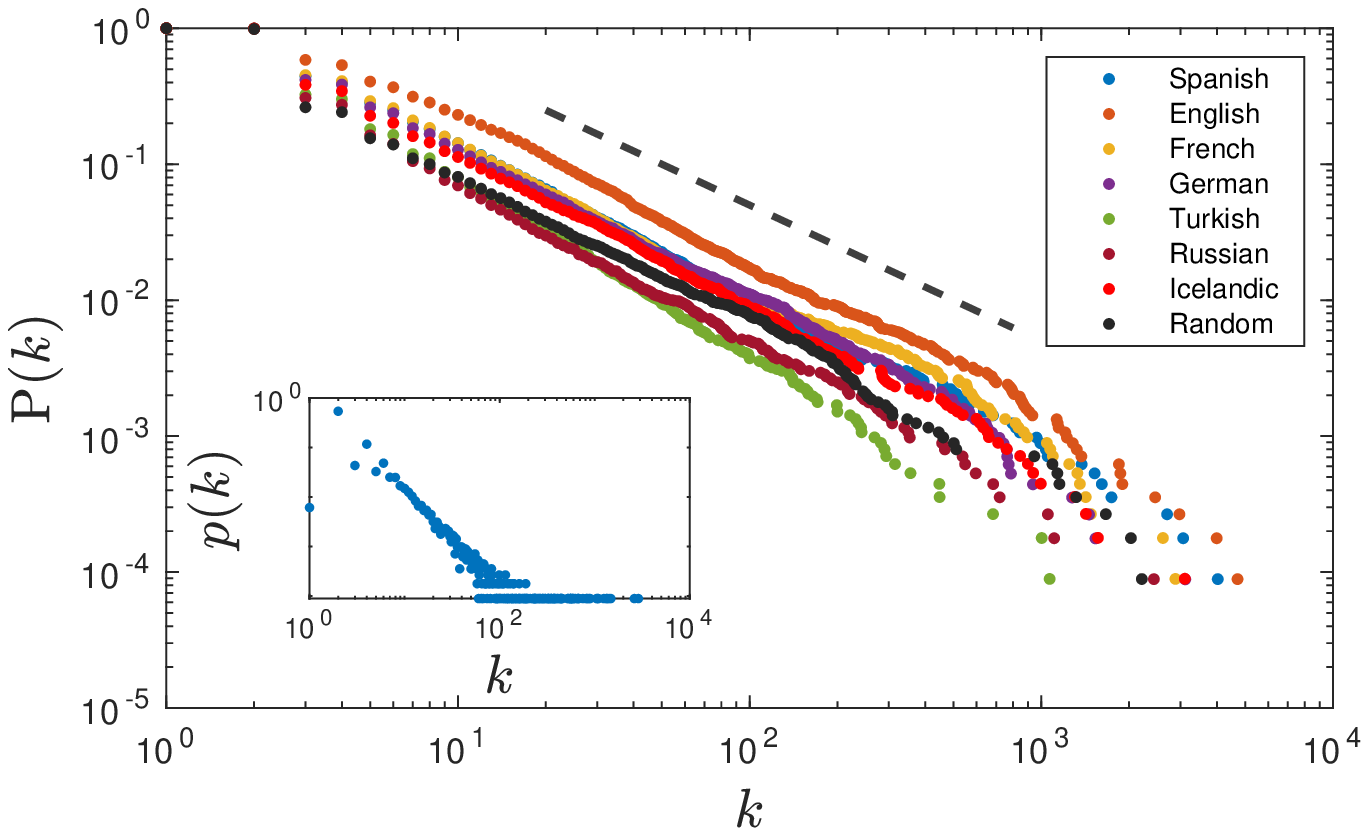}
\caption{\emph{Cumulative of the degree distribution $\mathrm{P}(f)\equiv\int_f^\infty \mathrm{p}(\zeta) d\zeta$ for single randomly chosen texts in each language: Spanish, English, French, German, Turkish, Russian, and Icelandic (colored symbols). As in figure 2, black dots represent a random text, and the dashed line corresponds to the behavior when the exponent in Zipf's law is $\alpha=1$. Inset: Degree distribution of the Don Quixote in French, note that the first few odd degrees $k=1,3,5,7$ deviate from the power law behavior.}}
  \label{degdist}
\end{figure}

Continuing with the universal laws describing texts, in figure (\ref{degdist}) we show an example of the degree distribution for the adjacency network of the texts studied in this work. It is clear that except for 
the low odd degrees ($k=1,3,5,7$, see inset in fig.(\ref{degdist})), the distribution is well described by 
a power law. The parameters corresponding to the texts are given in  Appendix(\ref{Resultados}). As mentioned previously, this asymptotic behavior is a consequence of Zipf's law. If we assume 
that each time a word appears, the input degree $k_{in}$ (alternatively, 
the output degree $k_{out}$) of the corresponding node increases 
approximately by one, then the input degree could be expected to grow 
proportional to the frequency of each word. Further, in general we can 
expect that the total degree of a node to be $k\approx k_{in}+k_{out}\approx 
2k_{in}$ (clearly this is not always true: for example, a word can appear twice, 
being preceded both times by the same word and followed by different 
words each time, leading to a degree $k=3$). Then, up to multiplicative 
factors, we can apply the same argument as in Equation \ref{eq.1}
for $\mathrm{p}(k)$, the degree distribution of the network, instead of $p(f)$ From this 
equation it again follows that if $f(n)\sim 1/n^\alpha$, then $\mathrm{p}(k)\sim 
1/k^{1+1/\alpha}$, which is again in close agreement with what we observe.

\section{Clustering coefficient}

Thus far, our results confirm that the all our texts exhibit the expected universal statistics observed in natural languages. Actually, it could be argued that these laws may be "too universal", not being able to clearly distinguish texts written  in real languages from our random texts. Further, all these laws appear to be consequence of Zipf's law, and this law reflects only the frequency of words, not their order. Thus, all three laws would still hold if the words of the texts were randomly shuffled. Clearly, shuffling the words destroys whatever relations may exist between successive words in a text, depending on the language in which it was written. This relation between successive words is what conveys meaning to a text.  Thus, we expect that the clustering coefficient \cite{BarabasiBook} of the adjacency network of each text,(constructed using words as nodes and linking those that are adjacent in the text), which depends strongly on the local structure, will distinguish between random texts and real texts, and even between texts in different languages. 

\begin{figure}[h!]
\centering
\includegraphics[scale=0.4]{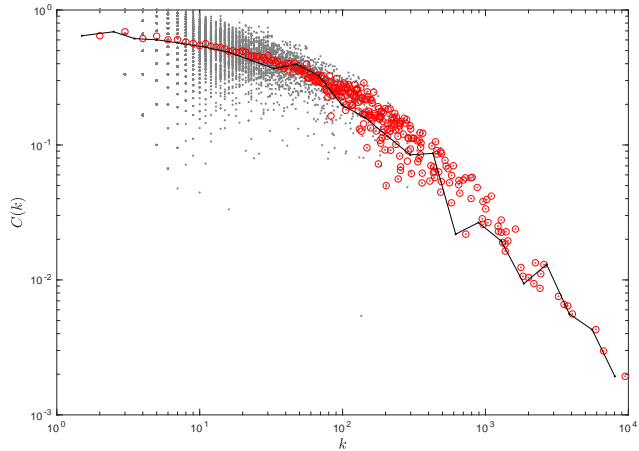}
\caption{\emph{Clustering coefficient as function of the degree for Don Quixote (Spanish). The gray dots represents the $C(k)$ for each node. Red circles are the average of $C(k)$. The black line is the logarithmic binning of the average.}}
  \label{C(k)}
\end{figure}

The clustering coefficient $C_i(k_i)$ of node $i$ with degree $k_i$ is 
defined as the ratio of the number of links between node $i$'s neighbors 
over the total number of links that would be possible for this node 
$k_i(k_i-1)/2$. Thus, clearly, $0\leq C_i(k_i)\leq 1$. Hapax legomena, 
for example, mostly correspond to nodes with degree $k=2$, thus their 
clustering coefficient can only take the values 0 and 1 (degree $k=1$ is 
possible if the hapax appears followed and preceded by the same word, 
but these are rare occurrences). In general terms, the actual values of 
the clustering coefficients vary as a function of the size of the 
network \cite{MeanFieldClustering}, thus, in order to compare the clustering coefficients of 
networks corresponding to different texts, we have trimmed our texts so 
they all have approximately the same vocabulary size ($\simeq 11260$). In figure (\ref{C(k)}) we show an example of the clustering coefficient as a function of 
$k$. There are many values $C(k)$ for each $k$ corresponding to the diverse nodes with the same degree. The red points in the graph denote the average clustering coefficient for each $k$, and the solid black line is the log-binning of this average.

\section{Language differentiation}

In order to quantify differences between languages, for each text we define the quantity 
$\nu(C)$ as
\begin{equation}\label{N(C)}
\nu(C) = \frac{\text{Number of nodes with same $C$ value }}{\text{Vocabulary}}
\end{equation}

\begin{figure}[h!]
\centering
\includegraphics[scale=0.7]{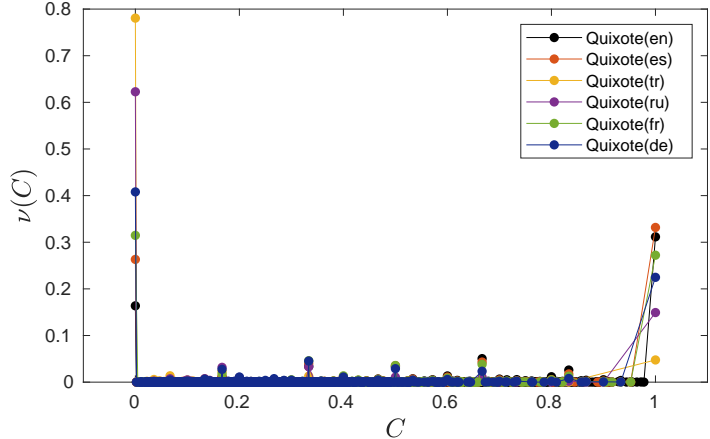}
\caption{\emph{Fraction of nodes with same Clustering Coefficient for Don Quixote in English, Spanish, Turkish, Russian, French and German. Note that nodes with $C=0\text{ and }1$ present the largest variability between different languages}}
  \label{NvsC}
\end{figure}
 In figure (\ref{NvsC}) we show $\nu(C)$ vs $C$ for Don Quixote in six different languages. From the graph it is clear that $\nu(0)$ and $\nu(1)$ show the largest degree of variation between the various languages, thus, we propose to focus on these two numbers to characterize the various languages.

\begin{figure}[h!]
\centering
\includegraphics[scale=0.5]{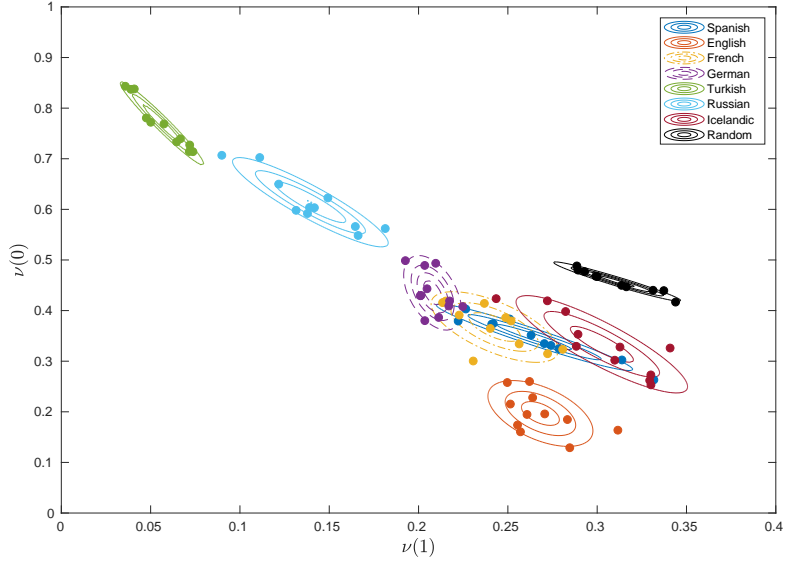}
\caption{\emph{Bi-variate normal distribution for $\nu(0)$ and $\nu(1)$ for the different texts and random sequences. Note that differences in the distributions are clear for languages that are known to be part of different linguistic families, for example Turkish and English. Languages that belongs to the same family (Spanish and French) are essentially indistinguishable.}}
  \label{galaxias}
\end{figure}

 In figure (\ref{galaxias}) we show a scatter plot of $\nu(1)$ vs $\nu(0)$ for the texts in every language presented here. Using maximum likelihood estimators, we fit correlated bi-variate Gaussian distributions to the scatter plots of each language, the contour plots of which are also shown in the graph. First and most importantly, we can see in the figure that there is a clear distinction
between languages and random texts. Also, we can see that languages 
tend to cluster in a way that is consistent with the known relationships among the languages. For example, in the figure we note that the contours corresponding to French and Spanish show a strong overlap, which might have 
been expected as they are closely related languages \cite{glo}. On the other 
hand, Russian is far from French and Spanish. This suggest that these 
curves may be used as a quantitative aid for the classification of languages into families. For example, French and Spanish which are both Romance languages, appear closer to each other than to Russian and Turkish, which have different origins.

In order to test the validity our results, we calculate $\nu(0)$ and $\nu(1)$ for 
another set of books, (see tables in the appendix (\ref{Libros})) and using the fitted Gaussian distributions for each language, we calculated the probability that a text in each language would have those values, which allows us to assign a likelihood that a text is written in one or another language.

\begin{table}[]
\scriptsize
\centering
\begin{tabular}{@{}lcccccccc@{}}
\toprule
Books                 & Spanish    & English      & French   & German     & Turkish    & Russian    & Icelandic   & Random      \\ \midrule
MobyDick(es)          &     10.572 &   0.00014223 &   125.25 &     9.1582 &         0 &    0.19125  &      4.0324 &         0 \\
TwentyThousand...(es) &     250.58 &    0.0033275 &   182.17 &     34.068 &         0 &   0.019141  &     0.45346 &         0 \\
TwentyYearsLater(en)  &          0 &       230.94 & 0.013046 &          0 &         0 &          0  &   1.179e-07 &         0 \\
BramStoker(en)        &          0 &       65.208 & 0.036547 &          0 &         0 &          0  &  0.00013916 &         0 \\
Voltaire(fr)          &     266.07 &     0.003546 &   196.17 &     11.899 &         0 &   0.022266  &     0.91734 &         0 \\
Miserables(fr)        &      23.99 &   0.00077475 &    127.1 &  0.0030604 &         0 &    0.02086  &       21.05 &         0 \\
MobyDick(de)          &   0.026313 &   7.5812e-07 &   31.707 &      325.5 &         0 &     2.6555  &      0.8309 &         0 \\
Dostoevsky(de)        &  0.0023808 &   0.00076044 &   4.8089 &     6.3034 &         0 & 2.4945e-07  &  1.9212e-06 &         0 \\
MobyDick(tr)          &          0 &            0 &        0 &          0 &    977.45 &     2.4039  &           0 &         0 \\
JulesVerne(tr)        &          0 &            0 &        0 &          0 &    25.009 &    0.77189  &           0 &         0 \\
AroundWorld...(ru)    &          0 &            0 &        0 &          0 & 0.0098057 &      53.27  &           0 &         0 \\
MysteriousIsland(ru)  &          0 &            0 &        0 &          0 &    5.7008 &     13.406  &           0 &         0 \\
Smásögur I(is)        &          0 &            0 & 0.022572 &          0 &         0 &          0  &      15.199 &         0 \\
Smásögur II(is)       &          0 &            0 &  0.58821 &          0 &         0 & 4.4908e-05  &      23.281 &         0 \\
RandomTextA           &          0 &            0 &        0 &          0 &         0 &          0  &  4.7682e-07 &    5.8013 \\
RandomTextB           &          0 &            0 &        0 &          0 &         0 &          0  &           0 &  0.045203 \\ \bottomrule \\
\end{tabular}
\caption{Probability density function for different texts written in several languages, and random texts. Values less than $1\times 10 ^{-8}$ are neglected.}
\label{tableII}
\end{table}

In table \ref{tableII} we can see, for example, that it is most likely that Smásögur I 
(Short stories in Icelandic) are written in Icelandic than in any of the 
languages analyzed, or that they are a random text.

Not surprisingly, it is not so easy to tell if Voltaire in French, is really
written in French or in Spanish, likewise, it is not easy to tell if Moby Dick
in Spanish is written in Spanish or French, and in both cases the maximum likelihood prediction fails. Nevertheless, it is clear that these books are not written in any of the other languages presented here, nor do they correspond to a random text. On the other hand, Twenty thousand leagues under the sea in Spanish and Les Miserables in French, are correctly identified, as well as all the other texts analyzed, including the random texts.

 To try to pinpoint the origin of the differentiation between different languages, we note that an inspection of the nodes with $C=0$ and $1$ reveals that they mainly consist of hapax legomena (as noted before, hapax legomena only have $C$ values of $0$ and $1$). To measure the relative importance of these words, we calculate the ratio of hapax legomena to the total number of words with $C=0$ and $1$, we call this number $\nu'_{H}(C)$.
 
\begin{table}[]
\centering
\begin{tabular}{@{}lcc@{}}
\toprule
BookName                 & $\nu'_{H}(0)$    & $\nu'_{H}(1)$    \\ \midrule
Don Quixote              & 0.92524 & 0.81389 \\
The Count of Montecristo & 0.91299 & 0.87284 \\
The Three Musketeers     & 0.9087  & 0.85338 \\
Jane Austen              & 0.93604 & 0.80736 \\
Celebrated Crimes        & 0.91249 & 0.85042 \\
Les Miserables           & 0.91795 & 0.87119 \\
Anna Karenina            & 0.91475 & 0.83565 \\
War And Peace            & 0.91263 & 0.8445  \\
Brothers Karamazov       & 0.9144  & 0.82616 \\
Oscar Wilde              & 0.85503 & 0.85562 \\
Charles Dickens          & 0.91206 & 0.85831 \\
Twenty Years Later       & 0.92125 & 0.84953 \\
Bram Stoker              & 0.92612 & 0.84925 \\ \bottomrule
\\
\end{tabular}
\caption{Fraction of hapax legomena with clustering coefficient equal to $0$ or $1$ for English texts}
\label{HapaxTab}
\end{table}
 
In Table \ref{HapaxTab}, we show the fraction of hapax legomena of the words with $C=0,1$ for several texts in English. A value close to $1$ indicates that most of the nodes that contribute to $\nu'_H(C)$ are words that appear only once in the document. This indicates that the local structure around those words, i.e, the way that they relate in the adjacency network, is particular to each language, and seems to be a key for language differentiation.

\begin{table}[]
\centering
\begin{tabular}{@{}lcc@{}}
\toprule
Language  & $\overline{\nu'_{H}(0)}$ & $\overline{\nu'_{H}(1)}$ \\ \midrule
Spanish   & 0.9002     & 0.9146     \\ \
English   & 0.9131     & 0.8452     \\
French    & 0.9023     & 0.9287     \\
German    & 0.9008     & 0.9440     \\
Turkish   & 0.8079     & 0.9797     \\
Russian   & 0.8676     & 0.9618     \\
Icelandic & 0.9148     & 0.8980     \\
Random    & 0.9600     & 0.9637     \\ \bottomrule \\
\end{tabular}
\caption{Average values of $\nu'_H(C)$ for Spanish, English, French, German, Turkish, Icelandic and Random texts.}
\label{tabavg}
\end{table}

In the Table \ref{tabavg} we see the average of $\nu'_H(C)$ for each of the languages studied here. Note that for example the values are clearly different for Spanish and Turkish, similar for Spanish and French, and very different for all languages and random.

\section{Conclusions}

Zipf's law is one of the most universal statistics of natural languages. However, it may be too universal. While it may not strictly apply to sequences of independent random symbols with random spacings \cite{ZipfCancho}, it appears to describe random texts that conserve most of the correlations between successive symbols, as accurately as it describes texts written in real languages. Further, Heaps-Herdan law and the degree distribution of the adjacency network, appear to be consequences of Zipf's law, and are, thus, as universal. 

In this work we studied 91 texts in seven different languages, as well as random texts constructed by randomizing the spacings between words without altering the order of the letters in the text. We find that they are all well described by the universal laws. However, we also found that the distribution of clustering coefficients of the networks of each text appears to vary from one language to another, and to distinguish random texts from real languages. The nodes that vary the most among the distributions of $C(k)$ are those for which $C(k)$ is equal to $0$ or $1$. We fit the scatter plot of these nodes to bivariate Gaussian distributions, which allows us to define the likelihood that a text is written in each given language. This method was very successful identifying the languages in which test were written, only failing to distinguish a couple of texts, confusing texts french and spanish, which have a strong overlap. In Table (\ref{tableII}) we present the evidence that we can use the statistics
of clustering coefficient to measure a sort of distance between languages.

Though hapax legomena account for most of the value $\nu(C)$ for $C=0$ and 1, we found that the fraction $\nu'_H(C)$ of hapax to other words is similar for French and Spanish, and different for Spanish and, say, Turkish. Further, $\nu'_H(C)$ is different between random texts and the languages we study. These observations might give some clue to the mechanism by which the clustering coefficient, and in particular the local structure around hapax legomena, helps to  differentiate languages.

Unlike the work presented by Gamallo et. al \cite{Gamallo}, which is Corpus-based, our work uses a relatively small amount of texts. Also as we can see in tables presented in Appendix (\ref{Resultados}), the length of the texts we use is not necessarily the length of the complete work. Texts were cut at the appropriate length for all of them to have approximately the same vocabulary ($\simeq 11260$). Thus, actual lengths ranged from $368076$ words for the Jane Austen books in English, to $26347$ words for the text we called Turkish I. This is important not only for computational reasons, it may also be important for studies of the relation between languages for which large corpora do not exist, something very common in the linguistic studies of the indigenous languages.
The method proposed in this work can be useful in such cases, as small texts trimmed to fill some appropriate vocabulary size is the only necessary ingredient.

\section{Acknowledgments}
Diego Espitia acknowledges financial support through a doctoral scholarship from Consejo Nacional de Ciencia y Tecnología (CONACyT).

\newpage
\begin{appendices}

\section{Tables and Results}\label{Resultados}

In this appendix we present tables of results for the data analyzed in this work. Here $\alpha_k$ and $\sigma_k$ represent the exponent and standard error of the power law for the degree distribution of the co-occurrence networks $p(k) \propto 1/k^{\alpha_k} $, for $k> k_{min}$, where $k_{min}$ is the smallest degree for which the power law holds. Similarly, $\alpha_Z$ and $\sigma_z$ represents the exponent and standard error of the distribution of frequencies $p(f)\propto 1/f^{\alpha_z}$; for $f > f_{min}$ where now $f_{min}$ is the smallest frequency for which the power law is satisfied. The values of the Heap's law $\beta$ and $\sigma_h$ were obtained via least square fitting.

For the estimation of the parameters we use the Maximum Likelihood Estimation (MLE) method for discerning and quantifying power-law behavior in empirical data \cite{PowerLaw}. The MLE works as follows: assuming that the data fits a power law, we estimate $\alpha$ via

\begin{equation}\label{MLE_PK}
\hat{\alpha}^* =  1 + N \left[\sum_{i=1}^{N}\ln \frac{x_i}{x^*_{min} - \frac{1}{2}}\right]^{-1},
\end{equation}

where $x_i > x_{min}^*$ for $i=1,...N$ and using as $x^*_{min}$ each element of the data set $\{x\}$. Then, using the Kolmogorov–Smirnov test we find the distance $D$ between the cumulative distribution of the data set and the cumulative distribution $P_{(x^*_{min},\alpha^*)}(x)$. From these set of distances, we find the value which minimizes $D$, this $x_{min}$, is the smallest data for which the power law holds, and can be used to determine the parameter of the power law $\hat{\alpha}$. In order to perform a goodness of the fit test, we construct $1000$ synthetic data, using the previous $\hat{\alpha}$  and $x_{min}$. Now we can count the fraction of the synthetic distances that are larger than the distance obtained from the data. This fraction is known as \emph{p-value} If this \emph{p-value}$>0.1$, then the difference between the data set and the model can be attributed to statistical fluctuations alone; if it is small, the model is not a plausible fit to the data.\cite{PowerLaw}

\begin{center}
\textbf{Spanish}
\begin{table*}[h!]
\scriptsize
\centering
\begin{tabular}{@{}lcc|cccc|cccc|cc@{}}
\toprule
Book Name                 & Length & Vocabulary & $\alpha_k$ & $\sigma_k$&$k_{min}$ & \emph{p-value}& $\alpha_Z$ & $\sigma_Z$&$f_{min}$ & \emph{p-value}   & $\beta$  & $\sigma_h$ \\ \midrule
The Count of Montecristo  & 92378  & 11275      & 2.15       & 0.03      &  9     & 0.439         & 1.87       & 0.01        &   1      & 0.640            & 0.781    & 0.002      \\
Don Quixote               & 113068 & 11277      & 2.11       & 0.03      & 10     & 0.206         & 1.84       & 0.01        &   1      & 0.119            & 0.810    & 0.002      \\
The Three Musketeers      & 106869 & 11242      & 2.10       & 0.03      & 11     & 0.669         & 1.86       & 0.01        &   1      & 0.203            & 0.746    & 0.002      \\
Unamuno                   & 104769 & 11219      & 2.05       & 0.03      & 10     & 0.602         & 1.89       & 0.01        &   1      & 0.107            & 0.765    & 0.002      \\
Valle-Inclan              & 76657  & 11252      & 2.24       & 0.03      &  8     & 0.532         & 2.04       & 0.02        &   5      & 0.331            & 0.780    & 0.002      \\
Concha Espina             & 60356  & 11226      & 2.33       & 0.04      &  9     & 0.190         & 2.12       & 0.03        &   4      & 0.445            & 0.814    & 0.001      \\
Angelina                  & 71434  & 11281      & 2.23       & 0.03      &  8     & 0.180         & 2.02       & 0.02        &   3      & 0.583            & 0.810    & 0.002      \\
Iliad                     & 91203  & 11275      & 2.19       & 0.03      &  8     & 0.658         & 1.96       & 0.02        &   4      & 0.419            & 0.799    & 0.002      \\
Odyssey                   & 92381  & 11290      & 2.18       & 0.02      &  6     & 0.289         & 1.96       & 0.02        &   4      & 0.510            & 0.797    & 0.002      \\
Pio Baroja                & 85227  & 11273      & 2.21       & 0.03      &  8     & 0.601         & 2.03       & 0.03        &   7      & 0.362            & 0.787    & 0.001      \\
The White Company         & 76186  & 11232      & 2.18       & 0.03      &  9     & 0.126         & 1.97       & 0.02        &   3      & 0.510            & 0.786    & 0.002      \\
Moby Dick                 & 69986  & 11230      & 2.15       & 0.03      &  9     & 0.249         & 2.00       & 0.01        &   2      & 0.533            & 0.795    & 0.002      \\
TwentyThousand            & 76443  & 11214      & 2.22       & 0.03      &  9     & 0.105         & 2.01       & 0.02        &   3      & 0.860            & 0.788    & 0.001      \\ \bottomrule
\end{tabular}                                                                                                  
\label{ES_Table}
\end{table*}

\textbf{English}

\begin{table*}[h!]
\scriptsize
\centering
\begin{tabular}{@{}lcc|cccc|cccc|cc@{}}
\toprule
BookName                 & Length & Vocabulary & $\alpha_k$ & $\sigma_k$&$k_{min}$ &\emph{p-value}  & $\alpha_Z$ & $\sigma_Z$&$f_{min}$&\emph{p-value}  & $\beta$     & $\sigma_h$ \\ \midrule
Don Quixote              & 221474 & 11278      & 2.16       & 0.03      & 17       & 0.417          & 1.90       & 0.02      & 15      &  0.823         & 0.731       & 0.003     \\
The Count of Montecristo & 178516 & 11261      & 2.17       & 0.03      & 19       & 0.244          & 1.97       & 0.03      & 16      &  0.934         & 0.703       & 0.002     \\
The Three Musketeers     & 233220 & 11266      & 2.14       & 0.03      & 24       & 0.972          & 1.91       & 0.03      & 33      &  0.579         & 0.704       & 0.003     \\
Jane Austen              & 368076 & 11270      & 2.16       & 0.03      & 34       & 0.805          & 1.93       & 0.04      & 84      &  0.702         & 0.660       & 0.003     \\
Celebrated Crimes        & 156044 & 11274      & 2.20       & 0.03      & 17       & 0.569          & 2.05       & 0.04      & 28      &  0.505         & 0.726       & 0.002     \\
Les Miserables           & 131649 & 11254      & 2.17       & 0.04      & 19       & 0.198          & 1.97       & 0.03      & 10      &  0.867         & 0.736       & 0.002     \\
Anna Karenina            & 259749 & 11268      & 2.13       & 0.03      & 20       & 0.965          & 1.86       & 0.02      & 11      &  0.105         & 0.687       & 0.002     \\
War And Peace            & 201580 & 11223      & 2.17       & 0.04      & 33       & 0.612          & 1.94       & 0.03      & 25      &  0.590         & 0.699       & 0.002     \\
Brothers Karamazov       & 291642 & 11212      & 2.12       & 0.03      & 27       & 0.647          & 1.85       & 0.02      & 22      &  0.600         & 0.686       & 0.003     \\
Oscar Wilde              & 174912 & 11262      & 2.15       & 0.04      & 28       & 0.865          & 1.92       & 0.03      & 24      &  0.774         & 0.716       & 0.002     \\
Charles Dickens          & 183844 & 11266      & 2.12       & 0.03      & 20       & 0.738          & 1.89       & 0.02      &  9      &  0.992         & 0.714       & 0.002     \\
Twenty Years Later       & 231543 & 11257      & 2.12       & 0.04      & 29       & 0.854          & 1.92       & 0.04      & 44      &  0.718         & 0.701       & 0.003     \\
Bram Stoker              & 221752 & 11265      & 2.13       & 0.03      & 23       & 0.182          & 1.88       & 0.03      & 20      &  0.804         & 0.691       & 0.002     \\ \bottomrule
\end{tabular}
\label{EN_Table}
\end{table*}

\newpage
\textbf{French}

\begin{table*}[h!]
\scriptsize
\centering
\begin{tabular}{@{}lcc|cccc|cccc|cc@{}}
\toprule
Book Name                             & Length & Vocabulary & $\alpha_k$   & $\sigma_k$&$k_{min}$&\emph{p-value}   & $\alpha_Z$   & $\sigma_Z$&$f_{min}$&\emph{p-value}  & $\beta$ & $\sigma_h$    \\ \midrule
The Count of Montecristo              & 105525 & 11271      & 2.10         & 0.03      &  9      & 0.378           & 1.89         & 0.02      &      3  & 0.681       & 0.745      & 0.002 \\
Don Quixote                           & 111728 & 11237      & 2.10         & 0.02      &  8      & 0.495           & 1.89         & 0.02      &      5  & 0.628       & 0.746      & 0.002 \\
The Three Musketeers                  & 111274 & 11268      & 2.07         & 0.03      & 11      & 0.520           & 1.85         & 0.01      &      1  & 0.326       & 0.768      & 0.002 \\
Oscar Wilde                           & 85015  & 11206      & 2.15         & 0.03      &  8      & 0.422           & 1.92         & 0.01      &      1  & 0.538       & 0.783      & 0.002 \\
Madame Bobary                         & 72966  & 11292      & 2.22         & 0.03      &  8      & 0.001           & 2.00         & 0.01      &      2  & 0.940       & 0.782      & 0.002 \\
Honoré de Balzac                      & 78495  & 11264      & 2.17         & 0.03      &  9      & 0.062           & 1.98         & 0.02      &      3  & 0.160       & 0.799      & 0.002 \\
Homero                                & 149951 & 11236      & 2.11         & 0.03      & 11      & 0.682           & 1.86         & 0.02      &      7  & 0.212       & 0.722      & 0.002 \\
Notre Dame                            & 69988  & 11282      & 2.18         & 0.03      &  9      & 0.012           & 1.98         & 0.01      &      1  & 0.965       & 0.784      & 0.001 \\
Lesuieur                              & 85886  & 11250      & 2.17         & 0.03      &  8      & 0.122           & 1.97         & 0.02      &      4  & 0.729       & 0.778      & 0.002 \\
Guy de Maupassant                     & 74709  & 11257      & 2.16         & 0.03      &  9      & 0.068           & 1.93         & 0.01      &      1  & 0.499       & 0.795      & 0.002 \\
Twenty Thousand Leagues Under the Sea & 74369  & 11272      & 2.23         & 0.03      &  8      & 0.001           & 2.00         & 0.02      &      3  & 0.895       & 0.781      & 0.002 \\
Voltaire                              & 81450  & 11267      & 2.15         & 0.03      &  9      & 0.002           & 1.95         & 0.01      &      2  & 0.160       & 0.772      & 0.001 \\
Les Miserables                        & 79011  & 11275      & 2.14         & 0.03      &  9      & 0.238           & 1.95         & 0.01      &      1  & 0.769       & 0.784      & 0.001 \\ \bottomrule
\end{tabular}
\label{FR_Table}
\end{table*}

\textbf{German}

\begin{table*}[h!]
 \scriptsize
\centering
\begin{tabular}{@{}lcc|cccc|cccc|cc@{}}
\toprule
Book Name                & Length & Vocabulary & $\alpha_k$ & $\sigma_k$&$k_{min}$&\emph{p-value} & $\alpha_Z$ & $\sigma_Z$&$f_{min}$&\emph{p-value}  & $\beta$ & $\sigma_h$        \\ \midrule
The Count of Montecristo & 99693  & 11263      & 2.06       & 0.02      &      8  & 0.669         & 1.82       & 0.01      &       1 & 0.156          & 0.738      & 0.002 \\
Don Quixote              & 81741  & 11323      & 2.07       & 0.03      &     10  & 0.716         & 1.92       & 0.01      &       1 & 0.385          & 0.921      & 0.006 \\
The Three Musketeers     & 107870 & 11271      & 2.04       & 0.03      &     13  & 0.623         & 1.82       & 0.01      &       1 & 0.629          & 0.743      & 0.002 \\
Honoré de Balzac         & 75986  & 11287      & 2.05       & 0.03      &     11  & 0.414         & 1.93       & 0.01      &       1 & 0.772          & 0.783      & 0.002 \\
Rudolf Hans Bartsch      & 58874  & 11288      & 2.07       & 0.04      &     18  & 0.496         & 1.94       & 0.03      &       5 & 0.705          & 0.805      & 0.002 \\
Felix Dahn I             & 67330  & 11268      & 2.17       & 0.05      &     23  & 0.616         & 1.96       & 0.01      &       1 & 0.404          & 0.785      & 0.002 \\
Felix Dahn II            & 75792  & 11257      & 2.09       & 0.02      &      8  & 0.658         & 1.91       & 0.01      &       1 & 0.248          & 0.781      & 0.002 \\
Charles Dickens I        & 82374  & 11274      & 2.06       & 0.02      &      8  & 0.128         & 1.90       & 0.01      &       1 & 0.853          & 0.779      & 0.002 \\
Cahrles Dickens II       & 81893  & 11285      & 2.00       & 0.03      &      9  & 0.256         & 1.92       & 0.01      &       1 & 0.536          & 0.822      & 0.003 \\
Alfred Döblin            & 56757  & 11240      & 2.12       & 0.03      &     10  & 0.595         & 2.02       & 0.02      &       2 & 0.526          & 0.787      & 0.002 \\
Gustave Falke            & 62815  & 11202      & 2.07       & 0.03      &     10  & 0.939         & 1.96       & 0.02      &       3 & 0.225          & 0.788      & 0.002 \\
MobyDick                 & 72414  & 11215      & 2.08       & 0.03      &     12  & 0.676         & 1.94       & 0.01      &       2 & 0.329          & 0.779      & 0.002 \\
Crime and Punishment     & 96492  & 11260      & 2.19       & 0.05      &     28  & 0.366         & 1.80       & 0.01      &       1 & 0.879          & 0.756      & 0.002 \\ \bottomrule
\end{tabular}
\label{DE_Table}
\end{table*}

\textbf{Turkish}

\begin{table*}[h!]
\scriptsize
\centering
\begin{tabular}{@{}lcc|cccc|cccc|cc@{}}
\toprule
BookName                 & Length & Vocabulary & $\alpha_k$ & $\sigma_k$&$k_{min}$&\emph{p-value}& $\alpha_Z$ & $\sigma_Z$&$f_{min}$&\emph{p-value} & $\beta$ & $\sigma_h$ \\ \midrule
The Count of Montecristo & 42040  & 11198      & 2.26       & 0.06      &      19 & 0.524        & 2.07       & 0.02      &       2 & 0.455         & 0.822      & 0.001 \\
Don Quixote              & 35207  & 11241      & 2.27       & 0.05      &      12 & 0.162        & 2.18       & 0.01      &       1 & 0.310         & 0.881      & 0.002 \\
The Three Musketeers     & 40731  & 11280      & 2.22       & 0.04      &      10 & 0.145        & 2.07       & 0.02      &       2 & 0.317         & 0.857      & 0.002 \\
Tale of Two Cities       & 37838  & 11292      & 2.26       & 0.04      &      11 & 0.371        & 2.14       & 0.01      &       1 & 0.113         & 0.855      & 0.002 \\
Oscar Wilde              & 35065  & 11205      & 2.29       & 0.05      &      14 & 0.182        & 2.13       & 0.02      &       2 & 0.367         & 0.866      & 0.002 \\
Jules Verne I            & 35595  & 11264      & 2.29       & 0.05      &      12 & 0.713        & 2.11       & 0.02      &       2 & 0.592         & 0.845      & 0.001 \\
David Copperfield        & 39672  & 11213      & 2.23       & 0.04      &      10 & 0.711        & 2.09       & 0.02      &       2 & 0.595         & 0.854      & 0.002 \\
Crime and Punishment     & 39716  & 11279      & 2.25       & 0.04      &      12 & 0.197        & 2.10       & 0.01      &       1 & 0.756         & 0.855      & 0.002 \\
Turkish I                & 26347  & 11240      & 2.50       & 0.08      &      16 & 0.944        & 2.34       & 0.01      &       1 & 0.357         & 0.913      & 0.001 \\
Turkish II               & 26765  & 11244      & 2.43       & 0.07      &      14 & 0.934        & 2.31       & 0.02      &       2 & 0.204         & 0.904      & 0.002 \\
Turkish III              & 27564  & 11288      & 2.34       & 0.06      &      12 & 0.487        & 2.21       & 0.04      &       4 & 0.336         & 0.883      & 0.001 \\
MobyDick                 & 33500  & 11224      & 2.31       & 0.06      &      16 & 0.352        & 2.19       & 0.01      &       1 & 0.455         & 0.881      & 0.001 \\
Jules Verne II           & 40060  & 11225      & 2.25       & 0.04      &      10 & 0.386        & 2.06       & 0.01      &       1 & 0.189         & 0.863      & 0.002 \\ \bottomrule
\end{tabular}
\label{TR_Table}
\end{table*}

\newpage

\textbf{Russian}

\begin{table*}[h!]
\scriptsize
\centering
\begin{tabular}{@{}lcc|cccc|cccc|cc@{}}
\toprule
Book Name                             & Length & Vocabulary & $\alpha_k$ & $\sigma_k$&$k_{min}$&\emph{p-value}& $\alpha_Z$   & $\sigma_Z$&$f_{min}$&\emph{p-value}  & $\beta$ & $\sigma_h$ \\ \midrule
Don Quixote                           & 41169  & 11277      & 2.16       & 0.04      &      10 & 0.799        & 2.21         & 0.01      &      1  &  0.485         & 0.864      & 0.002 \\
The Count of Montecristo              & 47282  & 11234      & 2.16       & 0.04      &      10 & 0.615        & 2.11         & 0.01      &      1  &  0.367         & 0.802      & 0.002 \\
The Three Musketeers                  & 51306  & 11277      & 2.14       & 0.03      &      10 & 0.869        & 2.06         & 0.01      &      1  &  0.196         & 0.818      & 0.002 \\
Anna Karenina                         & 53333  & 11242      & 2.12       & 0.04      &      11 & 0.625        & 2.02         & 0.02      &      2  &  0.175         & 0.823      & 0.002 \\
War And Peace                         & 45596  & 11321      & 2.14       & 0.03      &       9 & 0.019        & 2.09         & 0.01      &      1  &  0.232         & 0.821      & 0.002 \\
Brothers Karamazov                    & 47083  & 11293      & 2.11       & 0.05      &      16 & 0.861        & 2.16         & 0.01      &      1  &  0.785         & 0.835      & 0.002 \\
Twenty Thousand Leagues Under the Sea & 35961  & 11297      & 2.29       & 0.05      &      10 & 0.766        & 2.21         & 0.01      &      1  &  0.108         & 0.865      & 0.002 \\
Anton Chekhov                         & 45423  & 11282      & 2.18       & 0.04      &      11 & 0.713        & 2.13         & 0.01      &      1  &  0.714         & 0.869      & 0.002 \\
Oscar Wilde                           & 43504  & 11321      & 2.10       & 0.04      &      12 & 0.792        & 2.00         & 0.04      &      7  &  0.624         & 0.823      & 0.002 \\
Honoré de Balzac                      & 35407  & 11280      & 2.15       & 0.05      &      12 & 0.886        & 2.05         & 0.04      &      5  &  0.429         & 0.881      & 0.002 \\
Twenty Years Later                    & 48539  & 11250      & 2.10       & 0.04      &      11 & 0.636        & 1.99         & 0.03      &      4  &  0.801         & 0.823      & 0.002 \\
Moby Dick                             & 34748  & 11234      & 2.16       & 0.05      &      11 & 0.578        & 2.07         & 0.03      &      4  &  0.856         & 0.857      & 0.002 \\
Crime and Punishment                  & 40035  & 11217      & 2.19       & 0.05      &      16 & 0.724        & 2.15         & 0.01      &      1  &  0.678         & 0.835      & 0.001 \\ \bottomrule
\end{tabular}
\label{RU_Table}
\end{table*}

\textbf{Icelandic}

\begin{table*}[h!]
\scriptsize
\centering
\begin{tabular}{@{}lcc|cccc|cccc|cc@{}}
\toprule
BookName         & Length & Vocabulary & $\alpha_k$ & $\sigma_k$&$k_{min}$&\emph{p-value}   & $\alpha_Z$ & $\sigma_Z$&$f_{min}$ &\emph{p-value}& $\beta$ & $\sigma_h$   \\ \midrule
TorfhildiHólm    & 73242  & 11202      & 2.18       & 0.06      &    29 & 0.838           & 1.97       & 0.01      &     1  & 0.156        & 0.773      & 0.002 \\
SagaI            & 99051  & 11184      & 2.03       & 0.02      &     8 & 0.148           & 1.88       & 0.01      &     1  & 0.569        & 0.753      & 0.001 \\
SagaII           & 141436 & 11248      & 1.95       & 0.02      &     6 & 0.501           & 1.76       & 0.01      &     1  & 0.551        & 0.714      & 0.002 \\
SagaIII          & 103020 & 11270      & 2.00       & 0.02      &     8 & 0.964           & 1.84       & 0.01      &     2  & 0.640        & 0.734      & 0.001 \\
SagaIV           & 116521 & 11235      & 1.99       & 0.02      &     6 & 0.256           & 1.81       & 0.01      &     1  & 0.102        & 0.735      & 0.002 \\
SagaV            & 106061 & 11290      & 1.98       & 0.02      &     6 & 0.465           & 1.84       & 0.01      &     1  & 0.659        & 0.729      & 0.001 \\
SagaVI           & 116956 & 11296      & 2.21       & 0.06      &    50 & 0.634           & 1.83       & 0.01      &     1  & 0.118        & 0.734      & 0.002 \\
SagaVII          & 119928 & 11287      & 2.20       & 0.06      &    49 & 0.794           & 1.81       & 0.01      &     1  & 0.216        & 0.742      & 0.001 \\
JónTrausti       & 66577  & 11238      & 2.05       & 0.03      &     9 & 0.278           & 1.94       & 0.02      &     3  & 0.553        & 0.785      & 0.001 \\
JónThoroddsen    & 89739  & 11249      & 2.02       & 0.03      &     8 & 0.148           & 1.85       & 0.02      &     4  & 0.273        & 0.757      & 0.001 \\
ÞorgilsGjallanda & 65357  & 11285      & 2.10       & 0.03      &     8 & 0.227           & 1.97       & 0.02      &     2  & 0.295        & 0.786      & 0.001 \\
SmásögurI        & 58932  & 11287      & 2.10       & 0.03      &     9 & 0.717           & 1.98       & 0.02      &     4  & 0.811        & 0.803      & 0.001 \\
SmásögurII       & 61272  & 11226      & 2.10       & 0.04      &    12 & 0.301           & 1.99       & 0.02      &     2  & 0.126        & 0.803      & 0.001 \\ \bottomrule
\end{tabular}
\label{IS_Table}
\end{table*}

\textbf{Random}
\begin{table*}[h!]
\scriptsize
\centering
\begin{tabular}{@{}lcc|cccc|cccc|cc@{}}
\toprule
Book Name   & Length & Vocabulary & $\alpha_k$ & $\sigma_k$&$k_{min}$&\emph{p-value}& $\alpha_Z$ & $\sigma_Z$&$f_{min}$&\emph{p-value}  & $\beta$  & $\sigma_h$   \\ \midrule
Random I    & 63904  & 11258      & 2.05       & 0.03      &     10  & 0.901        & 1.88       & 0.02      &      3  & 0.952          & 0.805    & 0.001 \\
Random II   & 62391  & 11251      & 2.00       & 0.04      &     12  & 0.335        & 1.88       & 0.02      &      3  & 0.678          & 0.788    & 0.001 \\
Random III  & 62619  & 11286      & 2.02       & 0.03      &      9  & 0.522        & 1.90       & 0.02      &      3  & 0.445          & 0.802    & 0.001 \\
Random IV   & 61148  & 11208      & 1.99       & 0.03      &     11  & 0.256        & 1.91       & 0.02      &      3  & 0.856          & 0.808    & 0.001 \\
Random V    & 63181  & 11291      & 2.04       & 0.03      &      8  & 0.407        & 1.93       & 0.02      &      2  & 0.225          & 0.791    & 0.001 \\
Random VI   & 62430  & 11302      & 2.00       & 0.04      &     14  & 0.294        & 1.87       & 0.03      &      5  & 0.247          & 0.796    & 0.001 \\
Random VII  & 66740  & 11224      & 2.10       & 0.06      &     29  & 0.588        & 1.88       & 0.04      &     10  & 0.704          & 0.804    & 0.001 \\
Random VIII & 65939  & 11251      & 1.98       & 0.03      &     10  & 0.008        & 1.86       & 0.02      &      4  & 0.478          & 0.812    & 0.002 \\
Random IX   & 62318  & 11247      & 2.03       & 0.03      &      9  & 0.258        & 1.90       & 0.02      &      3  & 0.151          & 0.810    & 0.001 \\
Random X    & 61574  & 11239      & 1.98       & 0.03      &     11  & 0.395        & 1.92       & 0.02      &      2  & 0.102          & 0.814    & 0.001 \\
Random A    & 66795  & 11277      & 2.01       & 0.03      &      9  & 0.812        & 1.87       & 0.03      &      5  & 0.523          & 0.797    & 0.001 \\
Random B    & 65996  & 11262      & 2.01       & 0.04      &     11  & 0.895        & 1.88       & 0.03      &      4  & 0.755          & 0.797    & 0.001 \\ \bottomrule
\end{tabular}
\label{RND_Table}
\end{table*}
\end{center}

\newpage

\section{Texts used }\label{Libros}
Here we present the text used in this work. The vast majority of the texts were obtained from the Gutemberg project, except for the texts in Russian, Turkish and Icelandic, which were obtained from other sources.
\begin{table}[!h]
\begin{minipage}{.5\linewidth}
\scriptsize
\begin{tabular}{@{}ll@{}}
\toprule
\multicolumn{2}{c}{{\bf Spanish}}                                                                                                                                                         \\ \midrule
Alexandre Dumas        & \begin{tabular}[c]{@{}l@{}}The Count of Montecristo \\ The Three Musketeers\end{tabular}                                                                   \\ \midrule
Miguel de Cervantes    & Don Quixote                                                                                                                                                \\ \midrule
Miguel de Unamuno      & \begin{tabular}[c]{@{}l@{}}Niebla\\ Una Historia De Pasión\end{tabular}                                                                                    \\ \midrule
Ramón del Valle-Inclan & \begin{tabular}[c]{@{}l@{}}Memorias Del Marqués De Bradomin: \\ Sonata De Otoño\\ Sonata De Verano\\ Sonata De Primavera\\ Sonata De Invierno\end{tabular} \\ \midrule
Concha Espina          & \begin{tabular}[c]{@{}l@{}}Agua De Nieve\\ La Esfinge Maragata \\ Dulce Nombre\end{tabular}                                                                \\ \midrule
Rafael Delgado         & Angelina                                                                                                                                                   \\ \midrule
Homer                  & \begin{tabular}[c]{@{}l@{}}Iliad\\ Odyssey\end{tabular}                                                                                                    \\ \midrule
Pío Baroja             & \begin{tabular}[c]{@{}l@{}}Memorias De Un Hombre De Acción: \\ El Aprendiz De Conspirador\\ Los Caminos Del Mundo\end{tabular}                             \\ \midrule
Arthur Conan Doyle     & The White Company                                                                                                                                          \\ \midrule
Herman Melville        & Moby Dick                                                                                                                                                  \\ \midrule
Jules Verne            & Twenty Thousand Leagues Under the Sea                                                                                                                      \\ \bottomrule
\end{tabular}
\medskip
 \caption{Source: Gutemberg Project}
    \label{español}
\end{minipage}\hfill
\begin{minipage}{.5\linewidth}
    \medskip
\scriptsize
\begin{tabular}{@{}ll@{}}
\toprule
\multicolumn{2}{c}{{\bf English}}                                                                                                                                                                                                  \\ \midrule
Miguel de Cervantes & Don Quixote                                                                                                                                                                                            \\  \midrule
Alexandre Dumas     & \begin{tabular}[c]{@{}l@{}}The Count of Montecristo \\ The Three Musketeers\\ Celebrated Crimes\\ Twenty Years Later\end{tabular}                                                                   \\ \midrule
Jane Austen         & \begin{tabular}[c]{@{}l@{}}Mansfield Park\\ Northanger Abbey\\ Persuasion\\ Sense and Sensibility\end{tabular}                                                                                         \\ \midrule 
Victor Hugo         & Les Miserables                                                                                                                                                                                         \\ \midrule
Leon Tolstói        & \begin{tabular}[c]{@{}l@{}}Anna Karenina\\ War and Peace\end{tabular}                                                                                                                                  \\ \midrule
Fyodor Dostoevsky   & Brothers Karamazov                                                                                                                                                                                     \\ \midrule
Oscar Wilde         & \begin{tabular}[c]{@{}l@{}}The Picture of Dorian Gray\\ The Happy Prince and Other Tales\\ De Profundis\\ A House Of Pomegranates\\ The Canterville Ghost\\ Selected Prose Of Oscar Wilde\end{tabular} \\ \midrule
Charles Dickens     & \begin{tabular}[c]{@{}l@{}}Oliver Twist\\ A Tale Of Two Cities\end{tabular}                                                                                                                            \\ \midrule
Bram Stoker         & \begin{tabular}[c]{@{}l@{}}Dracula\\ The Jewel of Seven Stars\end{tabular}                                                                                                                             \\ \bottomrule
\end{tabular}
\medskip
\caption{Source: Gutemberg Project}
\label{tab:second_table}
\end{minipage}
\end{table}


\begin{table}[!h]
\begin{minipage}{.5\linewidth}
    \medskip
\scriptsize
\begin{tabular}{@{}ll@{}}
\toprule
\multicolumn{2}{c}{\textbf{Turkish}}                                                                                                                                                                                    \\ \midrule
Alexandre Dumas                                                              & \begin{tabular}[c]{@{}l@{}}The Count of Montecristo\\ The Three Musketeers\end{tabular}                                                  \\ \midrule
Miguel de Cervantes                                                          & Don Quixote                                                                                                                              \\ \midrule
Charles Dickens                                                              & \begin{tabular}[c]{@{}l@{}}A Tale of Two Cities\\ David Copperfield\end{tabular}                                                         \\ \midrule
\begin{tabular}[c]{@{}l@{}}Turkish I\\ Turkish II\\ Turkish III\end{tabular} & \begin{tabular}[c]{@{}l@{}}Modern prose:\\ samples from literary texts and newspapers\end{tabular}                                       \\ \midrule
Jules Verne                                                                  & \begin{tabular}[c]{@{}l@{}}Twenty Thousand Leagues Under the Sea\\ From the Earth to the Moon\\ Around the World in 80 Days\end{tabular} \\ \midrule
Herman Melville                                                              & Moby Dick                                                                                                                                \\ \midrule
Fyodor Dostoevsky                                                            & Crime and Punishment                                                                                                                     \\ \bottomrule
\end{tabular}
\medskip
\caption{Source: www.ekitapcilar.com. \\ Turkish I, II and III were obtained from \\ University of Oxford Text Archive \\ (http://ota.ox.ac.uk/desc/0387)}
\label{turco}

\end{minipage}\hfill
\begin{minipage}{.5\linewidth}
    \medskip
\scriptsize
\begin{tabular}{@{}ll@{}}
\toprule
\multicolumn{2}{c}{\textbf{Russian}}                                                                                                                  \\ \midrule
Alexandre Dumas     & \begin{tabular}[c]{@{}l@{}}The Count of Montecristo\\ The Three Musketeers\\ Twenty Years Later\end{tabular}                    \\ \midrule
Miguel de Cervantes & Don Quixote                                                                                                                     \\ \midrule
Oscar Wilde         & \begin{tabular}[c]{@{}l@{}}The Portrait of Dorian Gray\\ De Profundis\end{tabular}                                              \\ \midrule
Honoré de Balzac    & \begin{tabular}[c]{@{}l@{}}Fater Goriot\\ A Woman of Thirty\end{tabular}                                                        \\ \midrule
Jules Verne         & \begin{tabular}[c]{@{}l@{}}Twenty Thousand Leagues Under the Sea\\ Mysterious Island\\ Around the World in 80 Days\end{tabular} \\ \midrule
Anton Chekhov       & Short Stories Compilation                                                                                                       \\ \midrule
Fyodor Dostoevsky   & Brothers Karamazov                                                                                                              \\ \midrule
Leo Tolstoy         & \begin{tabular}[c]{@{}l@{}}Anna Karenina\\ War And Peace\end{tabular}                                                           \\ \bottomrule
\end{tabular}
\medskip
\caption{Source: https://www.e-reading.club }
    \label{ruso}
\end{minipage}
\end{table}


\begin{table}[!h]
\begin{minipage}{.5\linewidth}

    \medskip
\scriptsize
\begin{tabular}{@{}ll@{}}
\toprule
\multicolumn{2}{c}{\textbf{French}}                                                                                                                                                                                                                                                                                                                                  \\ \midrule 
Miguel de Cervantes                                                       & Don Quixote                                                                                                                                                                                                                                                                              \\ \midrule
Alexandre Dumas                                                           & \begin{tabular}[c]{@{}l@{}}The Count of Montecristo \\ The Three Musketeers\end{tabular}                                                                                                                                                                                                 \\ \midrule
Victor Hugo                                                               & \begin{tabular}[c]{@{}l@{}}The Hunchback of Notre-Dame\\ Les Miserables\end{tabular}                                                                                                                                                                                                     \\ \midrule
Jules Verne                                                               & Twenty Thousand Leagues Under the Sea                                                                                                                                                                                                                                                     \\ \midrule
Guy de Maupassant                                                         & \begin{tabular}[c]{@{}l@{}}Ball of Fat\\ Moonlight\\ Contes de la Bécasse\end{tabular}                                                                                                                                                                                                   \\ \midrule
Oscar Wilde                                                               & \begin{tabular}[c]{@{}l@{}}The Portrait of Dorian Gray\\ Intentions\end{tabular}                                                                                                                                                                                                         \\ \midrule
Gustave Flaubert                                                          & Madame Bovary                                                                                                                                                                                                                                                                            \\ \midrule
Honoré de Balzac                                                          & \begin{tabular}[c]{@{}l@{}}The Human Comedy. Scenes from private life:\\ At the Sign of the Cat and Racket\\ The Ball at Sceaux\\ The Purse\\ The Vendetta\\ Madame Firmiani\\ A Second Home\\ Domestic Bliss \\ The Imaginary Mistress\\ Study of a Woman\\ Albert Savarus\end{tabular} \\ \midrule
Homer                                                                     & Iliad                                                                                                                                                                                                                                                                                    \\ \midrule
\begin{tabular}[c]{@{}l@{}}Daniel Lesueur\\ (Jeanne Lapauze)\end{tabular} & Amour D'Aujourd'Hui                                                                                                                                                                                                                                                    \\ \midrule
Voltaire                                                                  & Candide \\
\bottomrule
\end{tabular}
\medskip
    \caption{Source: Gutemberg Project}
    \label{frances}

\end{minipage}\hfill
\begin{minipage}{.5\linewidth}

    \medskip
\scriptsize
\begin{tabular}{@{}ll@{}}
\toprule
\multicolumn{2}{c}{\textbf{German}}                                                                                      \\ \midrule
Alexandre Dumas     & \begin{tabular}[c]{@{}l@{}}The Count of Montecristo\\ The Three Musketeers\end{tabular}            \\ \midrule 
Miguel de Cervantes & Don Quixote                                                                                        \\ \midrule
Honoré de Balzac    & \begin{tabular}[c]{@{}l@{}}Grosse Und Kleine Welt (Short Stories)\\ A Woman of Thirty\end{tabular} \\ \midrule
Rudolf Hans Bartsch & \begin{tabular}[c]{@{}l@{}}Grenzen der Menschheit\\ Vom sterbenden Rokoko\end{tabular}             \\ \midrule
Felix Dahn          & \begin{tabular}[c]{@{}l@{}}Ein Kampf um Rom I\\ Ein Kampf um Rom II\end{tabular}                   \\ \midrule
Charles Dickens     & \begin{tabular}[c]{@{}l@{}}Oliver Twist \\ A Tale of Two Cities\end{tabular}                       \\ \midrule
Alfred Döblin       & Die Lobensteiner reisen nach Böhmen                                                                \\ \midrule
Gustav Falke        & Der Mann im Nebel                                                                                  \\ \midrule
Herman Melville     & Moby Dick                                                                                          \\ \midrule
Fyodor Dostoevsky   & Crime and Punishment                                                                               \\ \bottomrule
\end{tabular}
    \caption{Source: Gutemberg Project}
    \label{aleman}
\end{minipage}
\end{table}


\newpage
\begin{tiny}
\begin{longtable}{@{}ll@{}}
\toprule
\multicolumn{2}{c}{\textbf{Icelandic}}                                                                                                                                                                                                                      \\ \midrule
Torfhildi Hólm    & Brynjólfur  Biskup  Sveinsson                                                                                                                                                                                                           \\ \midrule
Sagas I           & \begin{tabular}[c]{@{}l@{}}Bandamanna Saga\\ Bardar Saga\\ Bjarnar Saga\\ Droplaugarsona Saga\\ Gisla Saga\\ Hrafnkels Saga\\ Eiríks Saga\\ Eyrbyggja Saga\end{tabular}                                                                 \\ \midrule
Sagas II          & \begin{tabular}[c]{@{}l@{}}Brennu-Njáls Saga\\ Laxdæla Saga\end{tabular}                                                                                                                                                                \\ \midrule
Sagas III         & \begin{tabular}[c]{@{}l@{}}Egils Saga\\ Grettis Saga\end{tabular}                                                                                                                                                                       \\ \midrule
Sagas IV          & \begin{tabular}[c]{@{}l@{}}Finnboga Saga\\ Fljótsdæla Saga\\ Flóamanna Saga\\ Fóstbræðra Saga\\ Grænlendinga Saga\\ Gull-Þóris Saga\end{tabular}                                                                                        \\ \midrule
Sagas V           & \begin{tabular}[c]{@{}l@{}}Gunnars Saga\\ Gunnlaugs Saga\\ Hænsna-Þóris Saga\\ Hallfreðar Saga\\ Harðar Saga\\ Hávarðar Saga\\ Heiðarvíga Saga\\ Hrana Saga\end{tabular}                                                                \\ \midrule
Sagas VI          & \begin{tabular}[c]{@{}l@{}}Kjalnesinga Saga\\ Kormáks Saga\\ Króka-Refs Saga\\ Ljósvetninga Saga\\ Reykdæla Saga\\ Svarfdæla Saga\\ Þórðar Saga\end{tabular}                                                                            \\ \midrule
Sagas VII         & \begin{tabular}[c]{@{}l@{}}Þorsteins Saga Hvíta\\ Þorsteins Saga Síðu-Hallssonar\\ Valla-Ljóts Saga\\ Vatnsdæla Saga\\ Víga-Glúms Saga\\ Víglundar Saga\\ Vopnfirðinga Saga\\ Færeyinga Saga\\ Ölkofra Saga\\ Laxdæla Saga\end{tabular} \\ \midrule
Jón Trausti       & \begin{tabular}[c]{@{}l@{}}Anna  Frá  Stóruborg\\ Borgir\end{tabular}                                                                                                                                                                   \\ \midrule
Jón Thoroddsen    & Maður  Og Kona                                                                                                                                                                                                                          \\ \midrule
Þorgils Gjallanda & \begin{tabular}[c]{@{}l@{}}Upp  Við  Fossa\\ Gamalt  Og  Nýtt\end{tabular}                                                                                                                                                              \\ \midrule
Smásögur I        & \begin{tabular}[c]{@{}l@{}}Brúðardraugurinn\\ Írafells - Móri\\ Sagan  Af  Heljarslóðarorrustu\\ Ferðasaga\\ Þórðar  Saga  Geirmundarsonar\\ Grímur  Kaupmaður  Deyr\\ Hans  Vöggur\end{tabular}                                        \\ \midrule
Smásögur II       & \begin{tabular}[c]{@{}l@{}}Kærleiksheimilið\\ Brennivínshatturinn\\ Gulrætur\\ Í vinnunni\\ Einræða\\ Vordraumur\\ Kvöld, nótt, morgunn\end{tabular}\\ \bottomrule \\
\caption{\small{Source: All sagas were obtained from https://sagadb.org/.\\ The other texts were obtained from https://www.snerpa.is/net/index.html}}
\end{longtable}
\end{tiny}

\end{appendices}

\bibliographystyle{unsrt} 

\newpage

\bibliography{Diego0.1}  

\begin{thebibliography}{10}

\bibitem{Zipf}
George Zipf.
\newblock {\em Human behavior and the principle of least effort: an
  introduction to human ecology}.
\newblock Addison-Wesley Press, Cambridge, MA, 1st edition, 1949.

\bibitem{GellMann}
Murray Gell~Mann and Merritt Ruhlen.
\newblock The origin and evolution of word order.
\newblock {\em Proceedings of the National Academy of Sciences of the United
  States of America}, 108:17290--5, 10 2011.

\bibitem{Konto}
I.~{Kontoyiannis}, P.~H. {Algoet}, Y.~M. {Suhov}, and A.~J. {Wyner}.
\newblock Nonparametric entropy estimation for stationary processes and random
  fields, with applications to english text.
\newblock {\em IEEE Transactions on Information Theory}, 44(3):1319--1327, May
  1998.

\bibitem{Kohler}
Reinhard Köhler.
\newblock Syntactic structures: Properties and interrelations.
\newblock {\em Journal of Quantitative Linguistics}, 6(1):46--57, 1999.

\bibitem{Zanette}
D.~H. Zanette.
\newblock Statistical patterns in written language.
\newblock {\em arXiv:1412.3336}, 2014.

\bibitem{AltmannStat}
Eduardo Altmann and Martin Gerlach.
\newblock Statistical laws in linguistics.
\newblock {\em arXiv: 1502.03296}, 02 2015.

\bibitem{Voynich}
Kevin~Knight Sravana~Reddy.
\newblock What we know about the voynich manuscript.
\newblock {\em Proceedings of the 5th ACL-HLT Workshop on Language Technology
  for Cultural Heritage, Social Sciences, and Humanities}, page 78–86, 2011.

\bibitem{VoyAmancio}
Diego~R. Amancio, Eduardo~G. Altmann, Diego Rybski, Osvaldo~N. Oliveira, Jr,
  and Luciano da~F. Costa.
\newblock Probing the statistical properties of unknown texts: Application to
  the voynich manuscript.
\newblock {\em PLOS ONE}, 8(7):1--10, 07 2013.

\bibitem{VoyMontemurro}
Marcelo~A. Montemurro and Damián~H. Zanette.
\newblock Keywords and co-occurrence patterns in the voynich manuscript: An
  information-theoretic analysis.
\newblock {\em PLOS ONE}, 8(6):1--9, 06 2013.

\bibitem{Copiale}
Kevin Knight, Beáta Megyesi, and Christiane Schaefer.
\newblock The secrets of the copiale cipher.
\newblock {\em Journal for Research into Freemasonry and Fraternalism}, 05
  2012.

\bibitem{Copiale2}
Kevin Knight, Be{\'a}ta Megyesi, and Christiane Schaefer.
\newblock The copiale cipher.
\newblock In {\em Proceedings of the 4th Workshop on Building and Using
  Comparable Corpora: Comparable Corpora and the Web}, pages 2--9, Portland,
  Oregon, June 2011. Association for Computational Linguistics.

\bibitem{Piantadosi2014}
Steven~T. Piantadosi.
\newblock Zipf's word frequency law in natural language: A critical review and
  future directions.
\newblock {\em Psychonomic Bulletin {\&} Review}, 21(5):1112--1130, Oct 2014.

\bibitem{Zipf4Phrases}
Jake Ryland~Williams, Paul~R. Lessard, Suma Desu, Eric~M. Clark, James~P.
  Bagrow, Christopher~M. Danforth, and Peter Sheridan~Dodds.
\newblock Zipf's law holds for phrases, not words.
\newblock {\em Scientific Reports}, 5:12209 EP --, Aug 2015.
\newblock Article.

\bibitem{ZipfEspanoles}
Isabel Moreno-Sánchez, Francesc Font-Clos, and Álvaro Corral.
\newblock Large-scale analysis of zipf’s law in english texts.
\newblock {\em PLOS ONE}, 11(1):1--19, 01 2016.

\bibitem{Newman2000}
Mark Newman.
\newblock The power of design.
\newblock {\em Nature}, 405(6785):412--413, 2000.

\bibitem{FerreriCancho2005}
R.~Ferrer~i Cancho.
\newblock The variation of zipf's law in human language.
\newblock {\em The European Physical Journal B - Condensed Matter and Complex
  Systems}, 44(2):249--257, Mar 2005.

\bibitem{ZornigRandomZipf}
Peter Zörnig.
\newblock Zipf's law for randomly generated frequencies: explicit tests for the
  goodness-of-fit.
\newblock {\em Journal of Statistical Computation and Simulation},
  85(11):2202--2213, 2015.

\bibitem{WentianLi}
W.~{Li}.
\newblock Random texts exhibit zipf's-law-like word frequency distribution.
\newblock {\em IEEE Transactions on Information Theory}, 38(6):1842--1845, Nov
  1992.

\bibitem{HerdanBook}
Gustav Herdan.
\newblock {\em Type-token mathematics}.
\newblock The Hague: Mouton, 1960.

\bibitem{HeapsBook}
Harold~Stanley Heaps.
\newblock {\em Information Retrieval: Computational and Theoretical Aspects}.
\newblock Academic Press, 1978.

\bibitem{HeapsBaeza}
Ricardo Baeza-Yates and Gonzalo Navarro.
\newblock Block addressing indices for approximate text retrieval.
\newblock {\em Proceedings of the sixth international conference on Information
  and knowledge management - CIKM 97}, pages 69--82, 2000.

\bibitem{egghe_2007}
Leo Egghe.
\newblock Untangling herdans law and heaps law: Mathematical and informetric
  arguments.
\newblock {\em Journal of the American Society for Information Science and
  Technology}, 58(5):702–709, 2007.

\bibitem{CanchoRedes}
Ramon Ferrer~i Cancho, Ricard~V. Sol\'e, and Reinhard K\"ohler.
\newblock Patterns in syntactic dependency networks.
\newblock {\em Phys. Rev. E}, 69:051915, May 2004.

\bibitem{AmancioRedes}
Camilo Akimushkin, Diego~Raphael Amancio, and Osvaldo~Novais Oliveira, Jr.
\newblock Text authorship identified using the dynamics of word co-occurrence
  networks.
\newblock {\em PLOS ONE}, 12(1):1--15, 01 2017.

\bibitem{Liu2013}
HaiTao Liu and Jin Cong.
\newblock Language clustering with word co-occurrence networks based on
  parallel texts.
\newblock {\em Chinese Science Bulletin}, 58(10):1139--1144, Apr 2013.

\bibitem{JaposKeyword}
Y.~Matsuo and M.~Ishizuka.
\newblock Keyword extraction from a single document using word co-occurrence
  statistical information.
\newblock {\em International Journal on Artificial Intelligence Tools},
  13(01):157--169, 2004.

\bibitem{BarabasiBook}
Albert-L\'aszl\'o Barab\'asi.
\newblock {\em Network Science}.
\newblock Cambridge University Press, 1st edition, 2016.

\bibitem{Gamallo}
Pablo Gamallo, José~Ramom Pichel, and Iñaki Alegria.
\newblock From language identification to language distance.
\newblock {\em Physica A: Statistical Mechanics and its Applications}, 484:152
  -- 162, 2017.

\bibitem{Barry}
Barry~R. Chiswick and Paul~W. Miller.
\newblock Linguistic distance: A quantitative measure of the distance between
  english and other languages.
\newblock {\em Journal of Multilingual and Multicultural Development},
  26(1):1--11, 2005.

\bibitem{PETRONI2010}
Filippo Petroni and Maurizio Serva.
\newblock Measures of lexical distance between languages.
\newblock {\em Physica A: Statistical Mechanics and its Applications},
  389(11):2280 -- 2283, 2010.

\bibitem{MeanFieldClustering}
Agata Fronczak, Piotr Fronczak, and Janusz~A. Ho\l{}yst.
\newblock Mean-field theory for clustering coefficients in barab\'asi-albert
  networks.
\newblock {\em Phys. Rev. E}, 68:046126, Oct 2003.

\bibitem{Miller}
George~A. Miller.
\newblock Some effects of intermittent silence.
\newblock {\em The American Journal of Psychology}, 70(2):311--314, 1957.

\bibitem{ZipfCancho}
Ramon Ferrer-i Cancho and Brita Elvev\aa~g.
\newblock Random texts do not exhibit the real zipf's law-like rank
  distribution.
\newblock {\em PLOS ONE}, 5(3):1--10, 03 2010.

\bibitem{ZipfTwo}
Ramon~Ferrer i~Cancho and Ricard~V. Solé.
\newblock Two regimes in the frequency of words and the origins of complex
  lexicons: Zipf’s law revisited.
\newblock {\em Journal of Quantitative Linguistics}, 8(3):165--173, 2001.

\bibitem{NAUMIS200884}
G.G. Naumis and G.~Cocho.
\newblock Tail universalities in rank distributions as an algebraic problem:
  The beta-like function.
\newblock {\em Physica A: Statistical Mechanics and its Applications},
  387(1):84 -- 96, 2008.

\bibitem{glo}
Haspelmath~Martin Hammarström~Harald, Forkel~Robert.
\newblock Glottolog 4.0.
\newblock Available at
  \url{https://glottolog.org/resource/languoid/id/roma1334} (2019/08/15).

\bibitem{PowerLaw}
Aaron Clauset, Cosma~Rohilla Shalizi, and M.~E.~J. Newman.
\newblock Power-law distributions in empirical data.
\newblock {\em SIAM Review}, 51(4):661–703, 2009.

\end{thebibliography}
\end{document}